\documentclass[aps,prd,amsmath,amssymb,showpacs,]{revtex4}
\usepackage{graphicx}
\begin{document}

\title{$ZZ\gamma$ and $Z\gamma\gamma$ couplings in
$\gamma e$ collision with polarized beams}

\author{S. Ata\u{g} }
\email[]{atag@science.ankara.edu.tr}
\affiliation{Department of Physics, Faculty of Sciences, 
Ankara University, 06100 Tandogan, Ankara, Turkey}

\author{\.{I}. \c{S}ahin}
\email[]{sahin@science.ankara.edu.tr}
\affiliation{Department of Physics, Faculty of Sciences,
Ankara University, 06100 Tandogan, Ankara, Turkey}

\begin{abstract}
The potential of $\gamma$e mode of linear $e^{+}e^{-}$
collider to probe $ZZ\gamma$ and $Z\gamma\gamma$
vertices is investigated
through the Z boson production 
from the procees $\gamma e\to Z e$.
Considering the longitudinal and transverse  
polarization states
of the Z boson and incoming polarized beams 
we find the  95\% C.L.
limits on the form factors  $h_{3}^{Z}$, $h_{4}^{Z}$, 
$h_{3}^{\gamma}$ and $h_{4}^{\gamma}$
with integrated luminosity 500$fb^{-1}$ and  
$\sqrt{s}=$0.5, 1, 1.5  TeV energies. 
It is shown that the polarization can
improve sensitivities by  factors 2-3 depending on the energy.
\end{abstract}


\maketitle

\section{Introduction}
Studies of the trilinear gauge boson couplings provide 
important tests of the standard model (SM) of electroweak 
interactions. The SM predicts no tree-level couplings 
between the Z boson and the photon. Deviation of the 
couplings from the expected values would indicate the 
existence of new physics beyond the SM. Therefore 
precision measurements of triple vector boson vertices
will be the crucial tests of the structure of the SM.

Possible anomalous $ZZ\gamma$ and $Z\gamma\gamma$
couplings must obey Lorentz and gauge invariance.
Within the formalism of Ref. \cite{berger} there
are eight anomalous coupling parameters
$h_{i}^{Z}$, $h_{i}^{\gamma}$ $(i=1,..,4)$
which are all zero in the standard model.
Here we are interested in CP-even couplings
that are proportional to $h_{3}^{V}$ and $h_{4}^{V}$
($V=Z, \gamma$).
Due to partial wave unitarity constraints
at high energies, an energy dependent
form factor ansatz can be cosidered:
\begin{eqnarray}
h_{i}^{V}(\hat{s})=
\frac{h_{i0}^{V}}{(1+\hat{s}/\Lambda^{2})^{3}}
~~;~~i=1,3 \\
h_{i}^{V}(\hat{s})=
\frac{h_{i0}^{V}}{(1+\hat{s}/\Lambda^{2})^{4}}
~~;~~i=2,4
\end{eqnarray}
In this work we assume that new physics scale $\Lambda$ is above
the collision energy $\sqrt{\hat{s}}$.

Therefore, CP conserving anomalous
$Z(p_{1})\gamma(p_{2})Z(p_{3})$
vertex function  can be written following the papers of
\cite{berger}:

\begin{eqnarray}
ig_{e}\Gamma_{Z\gamma Z}^{\alpha\beta\mu}(p_{1},p_{2},p_{3})
=ig_{e}\frac{p_{3}^{2}-p_{1}^{2}}{m_{Z}^{2}} \left[h_{3}^{Z}
\epsilon^{\mu\alpha\beta\rho}p_{2\rho}+\frac{h_{4}^{Z}}{m_{Z}^{2}}
p_{3}^{\alpha}\epsilon^{\mu\beta\rho\sigma}p_{3\rho}p_{2\sigma}
\right]
\end{eqnarray}
where $m_{Z}$ and $g_{e}$ are the Z-boson mass and charge of the
proton.
The $Z\gamma\gamma$ vertex function
can be obtained with the replacements:

\begin{eqnarray}
\frac{p_{3}^{2}-p_{1}^{2}}{m_{Z}^{2}}\to
\frac{p_{3}^{2}}{m_{Z}^{2}} ,\;\;\; h_{i}^{Z}\to h_{i}^{\gamma}
,\;\;\; i=3,4
\end{eqnarray}
The overall factor $p_{3}^{2}$ in the $Z\gamma\gamma$
vertex function originates from electromagnetic gauge invariance.
Due to Bose statistics  the $Z\gamma\gamma$ vertex vanishes identically if both
photons are on shell (Yang's theorem) \cite{yang}.

Previous limits on the $ZZ\gamma$ and $Z\gamma\gamma$
anomalous couplings have been provided by the Tevatron 
$|h_{3}^{Z}|<0.36$, $|h_{4}^{Z}|<0.05$, 
$|h_{3}^{\gamma}|<0.37$ and $|h_{4}^{\gamma}|<0.05$
\cite{d0} and LEP experiments $-0.50<h_{3}^{Z}<0.36$, 
$-0.12<h_{4}^{Z}<0.39$,
$-0.33<h_{3}^{\gamma}<0.01$ and 
$-0.02<h_{4}^{\gamma}<0.24$
\cite{l3} at 95\% C.L. .

Based on the analysis of ZZ production at the 
upgraded Fermilab Tevatron and the CERN Large Hadron 
Collider (LHC) achievable limits on the $ZZ\gamma$
couplings have been discussed \cite{baur}.

Research and development on linear $e^{+}e^{-}$
colliders at SLAC, DESY and KEK have been progressing
and the physics potential of these future machines
is under intensive study.
After  linear  colliders are constructed
$\gamma e$ and $\gamma\gamma$ modes with real photons
should be discussed and may work as complementary
to basic colliders \cite{akerlof,barklow}.
Real gamma beam is obtained by the Compton
backscattering of laser photons off linear electron
beam where most of the photons are produced at the
high energy region. Since the luminosities for $\gamma e$
and $\gamma\gamma$ collisions turn out
to be of the same order as the one for $e^{+}e^{-}$
collision \cite{ginzb1}, the cross sections for
photoproduction processes with real photons
are considerably larger  than virtual photon case.
Polarizability of real gamma beam is an additional
advantage for polarized beam experiments. In this
paper we examine the capability of $\gamma e$ mode
of LC to probe anomalous $ZZ\gamma$ and 
$Z\gamma\gamma$ couplings from Z boson production 
with polarized electron and gamma beams, assuming Z
polarization will be measured.

\section{Cross Sections}

\begin{figure}
\includegraphics{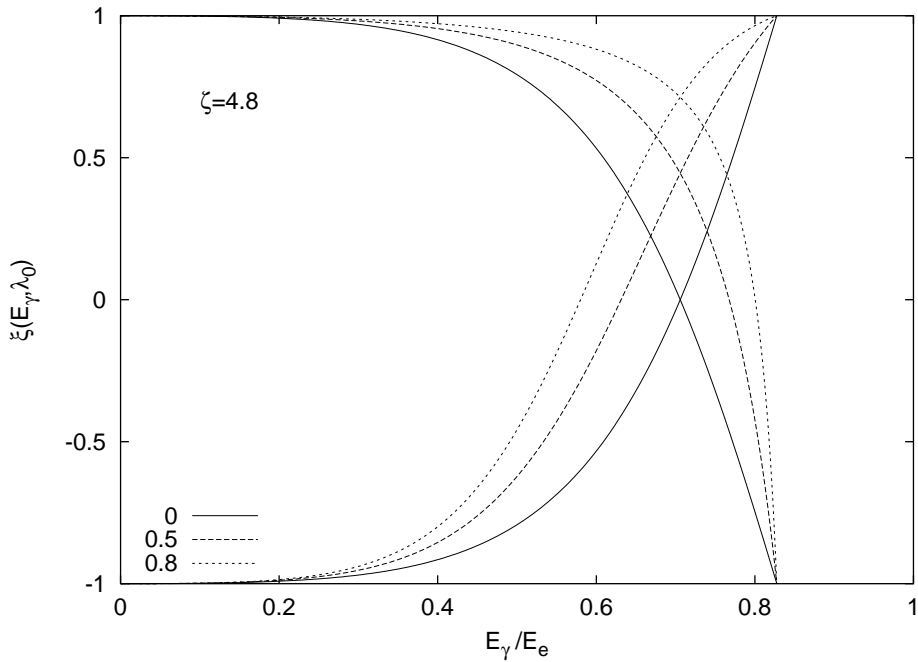}
\caption{Helicity of the backscattered photons as a function
of their energy. The set of curves from the bottom (upper)
are plotted for $\lambda_{0}=-1$ ($\lambda_{0}=1$) and the legends
are for helicities of the initial electron beam $\lambda_{e}$.
\label{fig1}}
\end{figure}

For $\gamma e\to Ze$ subprocess  
the helicity dependent differential cross section is given by

\begin{eqnarray}
\frac{d\hat{\sigma}(\lambda_{0},\lambda_{Z})}
 {d\cos\theta}=&&
\frac{\beta}{32\pi\hat{s}}
\{\frac{1}{4}(1-P_{e})
\left[(1+\xi(\omega , \lambda_{0}))|M(+,L;\lambda_{Z},L)|^{2}
+(1-\xi(\omega , \lambda_{0}))|M(-,L;\lambda_{Z},L)|^{2}\right]
 \nonumber \\
+&&\frac{1}{4}(1+P_{e})
\left[(1+\xi(\omega , \lambda_{0}))|M(+,R;\lambda_{Z},R)|^{2}
+(1-\xi(\omega , \lambda_{0}))|M(-,R;\lambda_{Z},R)|^{2}\right]
\}
\end{eqnarray}
where $\theta$ is the angle between incoming photon with 
helicity $\lambda_{\gamma}$ and outgoing Z boson with helicity 
$\lambda_{Z}$ in the c.m. frame.
Helicity amplitudes 
$M(\lambda_{\gamma}, \sigma_{e}; \lambda_{Z}, \sigma_{e}^{\prime})$
are given in the appendix. 
$\sigma_{e}$ and $\sigma_{e}^{\prime}$ are incoming and 
outgoing  electron helicities. 
Above cross section has been connected to initial laser photon helicity
$\lambda_{0}$ before Compton backscattering.
$P_{e}$ is the initial electron beam polarization and
$\xi(E_{\gamma},\lambda_{0})$ is the helicity of the Compton
backscattered photon \cite{ginzb1,telnov}

\begin{eqnarray}
\xi(E_{\gamma},\lambda_{0})={{\lambda_{0}(1-2r)
(1-y+1/(1-y))+\lambda_{e} r\zeta[1+(1-y)(1-2r)^{2}]}
\over{1-y+1/(1-y)-4r(1-r)-\lambda_{e}\lambda_{0}r\zeta
(2r-1)(2-y)}}
\end{eqnarray}
Here $r=y/[\zeta(1-y)]$, $y=E_{\gamma}/E_{e}$ and
$\zeta=4E_{e}E_{0}/M_{e}^{2}$. $E_{0}$ is the energy
of initial laser photon and $E_{e}$ and $\lambda_{e}$
are the energy and the helicity
of initial  electron beam before Compton backscattering.
One should note that $P_{e}$ and $\lambda_{e}$  refer to
different electron beams. In the cross section calculation 
we will keep $\lambda_{e}=0$ because of its minor contribution.
The behaviour of the helicity of backscattered
photons  can be observed from  Fig. \ref{fig1} as a function of
their energy.  From the figure we see that the backscattered
photons reach maximum polarization at highest energy region.
For outgoing Z bosons we take into account the possibility
that the transverse and
longitudinal polarizations can be observed for each
$\lambda_{0}$ state. 

The cross sections which
will be used in our calculations  are as follows

\begin{eqnarray}
{d\hat{\sigma}(\lambda_{0},TR)\over d\cos\theta}=
{d\hat{\sigma}(\lambda_{0},+)\over d\cos\theta}+
{d\hat{\sigma}(\lambda_{0},-)\over d\cos\theta}
\end{eqnarray}

\begin{eqnarray}
{d\hat{\sigma}(\lambda_{0},LO)\over d\cos\theta}=
{d\hat{\sigma}(\lambda_{0},0)\over d\cos\theta}
\end{eqnarray}
where TR stands for transverse and LO for longitudinal.
For the unpolarized beams the cross section takes the form
\begin{eqnarray}
{d\hat{\sigma}^{unpol}\over d\cos\theta}=
{d\hat{\sigma}(\lambda_{0},TR)\over d\cos\theta}+
{d\hat{\sigma}(\lambda_{0},LO)\over d\cos\theta}
\end{eqnarray}
with $\lambda_{0}=0$, $\lambda_{e}=0$ and $P_{e}=0$.

\begin{figure}
\includegraphics{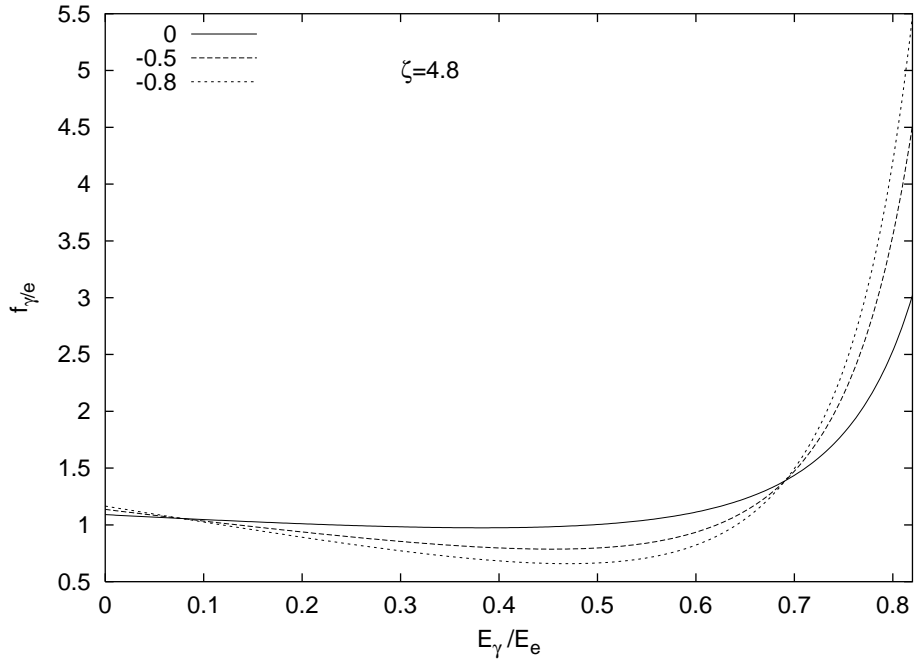}
\caption{Energy distribution of the backscattered photons for
$\lambda_{0}\lambda_{e}=0, -0.5, -0.8$.
\label{fig2}}
\end{figure}

For the integrated cross section, we need the spectrum of
backscattered photons in connection with helicities
of initial laser photon and  electron which
is given below \cite{ginzb1,telnov}

\begin{eqnarray}
f_{\gamma/e}(y)={{1}\over{g(\zeta)}}[1-y+{{1}\over{1-y}}
-{{4y}\over{\zeta(1-y)}}+{{4y^{2}}\over
{\zeta^{2}(1-y)^{2}}} + \lambda_{0}\lambda_{e} r\zeta
(1-2r)(2-y)]
\end{eqnarray}
where

\begin{eqnarray}
g(\zeta)=&&g_{1}(\zeta)+
\lambda_{0}\lambda_{e}g_{2}(\zeta) \nonumber\\
g_{1}(\zeta)=&&(1-{{4}\over{\zeta}}
-{{8}\over{\zeta^{2}}})\ln{(\zeta+1)}
+{{1}\over{2}}+{{8}\over{\zeta}}-{{1}\over{2(\zeta+1)^{2}}} \\
g_{2}(\zeta)=&&(1+{{2}\over{\zeta}})\ln{(\zeta+1)}
-{{5}\over{2}}+{{1}\over{\zeta+1}}-{{1}\over{2(\zeta+1)^{2}}}
\end{eqnarray}
The definitions of r, y and $\zeta$
are the same as in the helicity expression and the
maximum value of y reaches 0.83 when $\zeta=4.8$.
To see the influence of polarization,
energy distributions of backscattered photons $f_{\gamma/e}$
is plotted  for $\lambda_{0}\lambda_{e}$=0, -0.5, -0.8
in Fig. \ref{fig2}.

The expression of the integrated cross section over
the backscattered photon spectrum is written below for 
completeness:

\begin{eqnarray}
{d{\sigma}(\lambda_{0},\lambda_{Z})
\over d\cos\theta}=\int_{y_{min}}^{0.83}
f_{\gamma/e}(y){d\hat{\sigma}(\lambda_{0},\lambda_{Z})
\over d\cos\theta}dy
\end{eqnarray}
with $y_{min}=M_{Z}^{2}/s$. Here $\hat{s}$ is related to $s$,
the square of the center of mass energy of $e^{+}e^{-}$ system,
by $\hat{s}=ys$.

\begin{figure}
\includegraphics{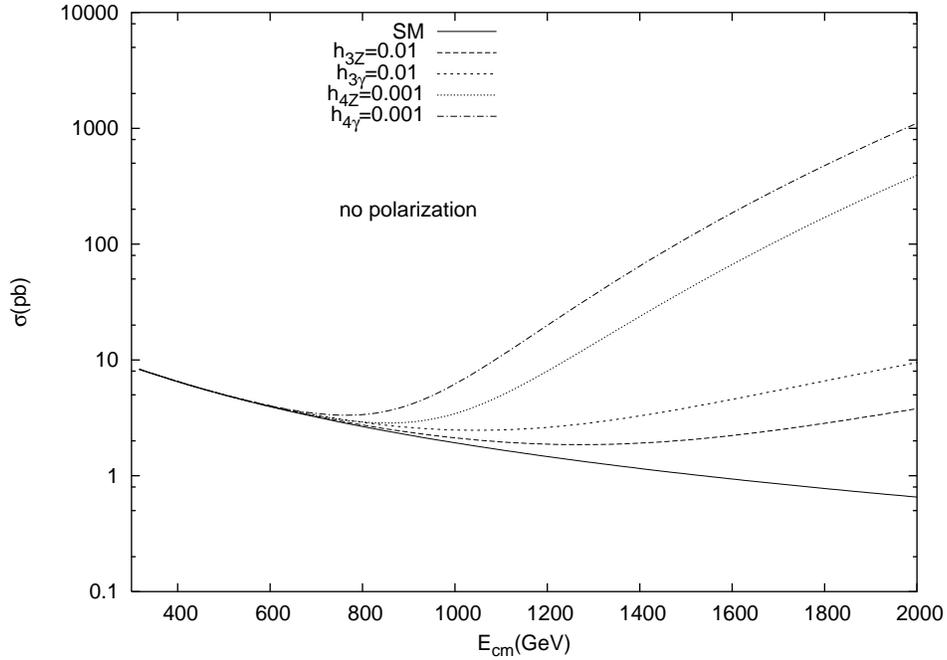}
\caption{Total cross section versus the center of mass energy 
$\sqrt{s}$ of the parental linear $e^{+}e^{-}$ 
colliders for the unpolarized
case. CP conserving coupling parameters of $ZZ\gamma$ 
and $Z\gamma\gamma$ vertices are shown on the figure. 
\label{fig3}}
\end{figure}

\begin{figure}
\includegraphics{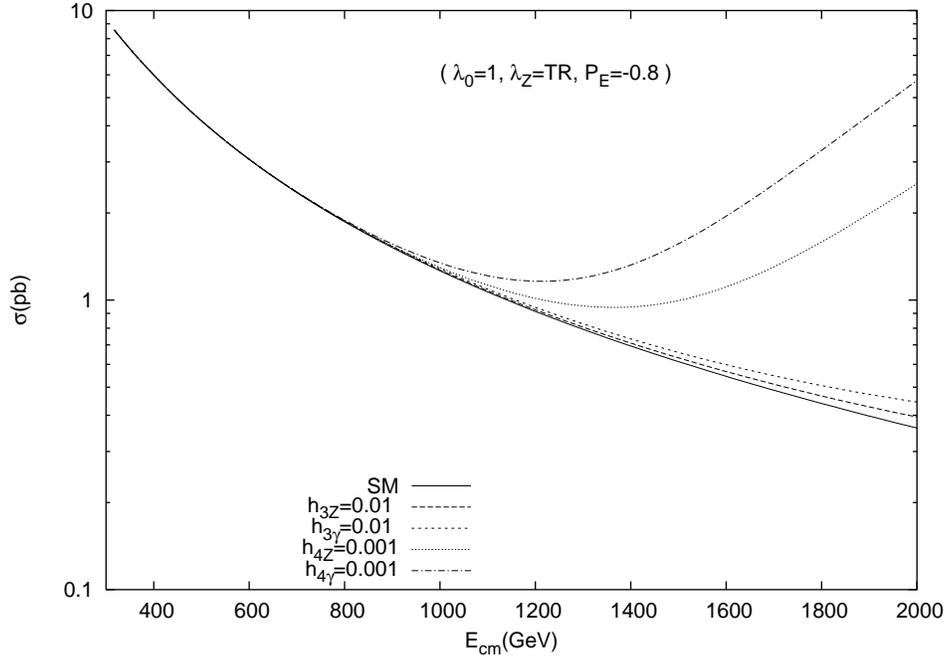}
\caption{The same as in Fig.\ref{fig3} but for polarization 
parameters $\lambda_{0}=1$, $\lambda_{Z}=TR$ and  $P_{E}=-0.8$. 
TR and LO are for transverse and longitudinal polarization. 
\label{fig4}}
\end{figure}

\begin{figure}
\includegraphics{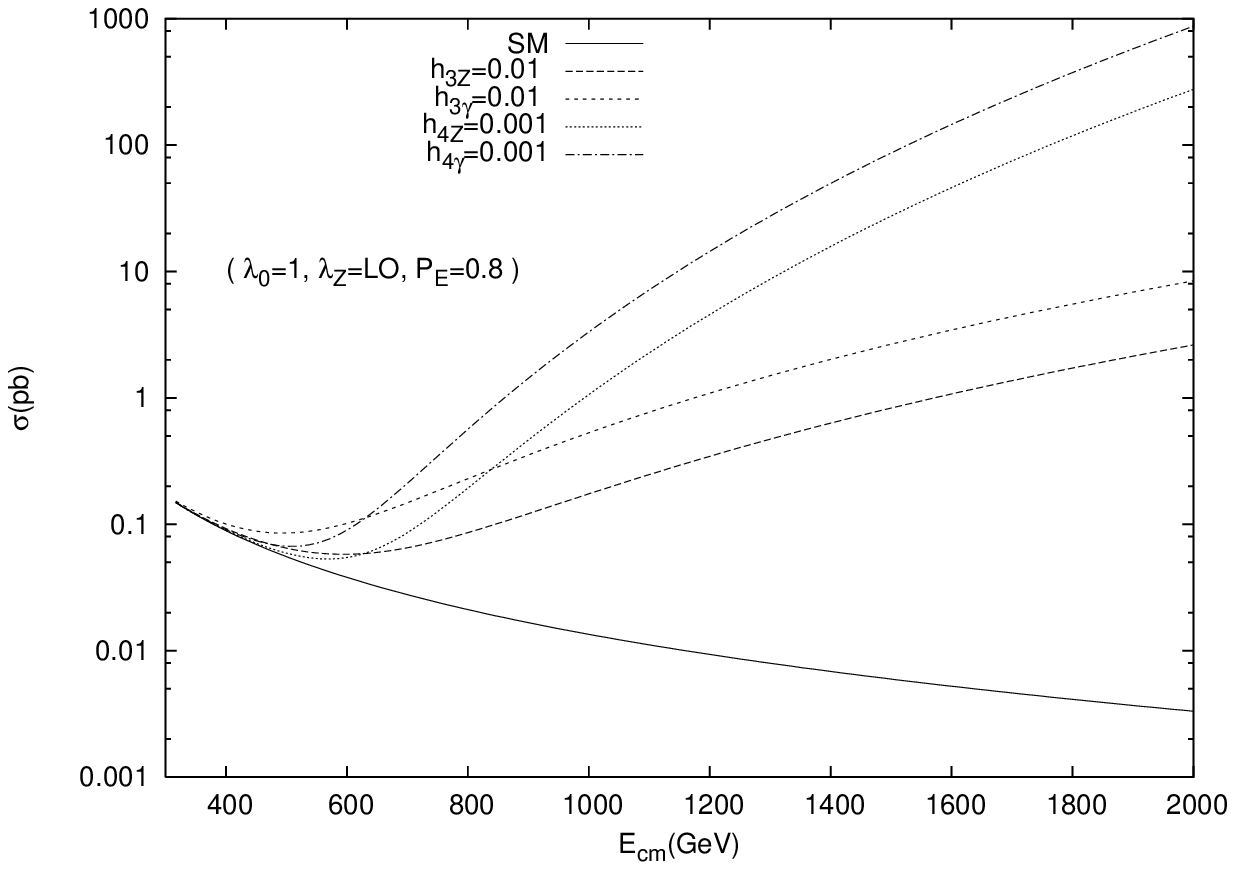}
\caption{The same as in Fig.\ref{fig3} but for polarization
parameters $\lambda_{0}=1$, $\lambda_{Z}=LO$ and  $P_{E}=0.8$.
\label{fig5}}
\end{figure}

In order to give an idea about the comparison
of unpolarized and polarized cases, integrated total cross
sections as functions of $\sqrt{s}$
are shown in Figs. \ref{fig3}-\ref{fig5} for various
coupling values and different configurations
of polarizations. From figures we see gobal enhancement 
of the cross sections when anomalous contributions 
are included especially after $\sqrt{s}\sim 1$ TeV region.  
Common feature in each figure is that
higher $\sqrt{s}$ will highly
improve the sensitivity of the cross section to
$h_{4}^{Z}$ and $h_{4}^{\gamma}$  when compared with 
$h_{3}^{Z}$ and $h_{3}^{\gamma}$.
Another important faeture is the longitudinal polarization 
of Z boson which  makes the separation much higher 
than the  standard model. 
As can be seen from Fig. \ref{fig1} 
$\xi(E_{\gamma},\lambda_{0})$ is antisymmetric with 
respect to $\lambda_{0}$ when $\lambda_{e}=0$. Thus, the 
contribution of the factor $\xi(E_{\gamma},\lambda_{0})P_{e}$
to the cross section does not change by reversing the 
sign of both $\lambda_{0}$ and $P_{e}$. This leads to the 
fact that we get almost the same plots as 
Figs. \ref{fig4} and \ref{fig5} if we reverse the the sign of both 
$\lambda_{0}$ and $P_{e}$. This feature also appears in the angular 
distributions below and on the tables of
sensitivities in the next section. 
This is why we avoid to plot more figures. 

\begin{figure}
\includegraphics{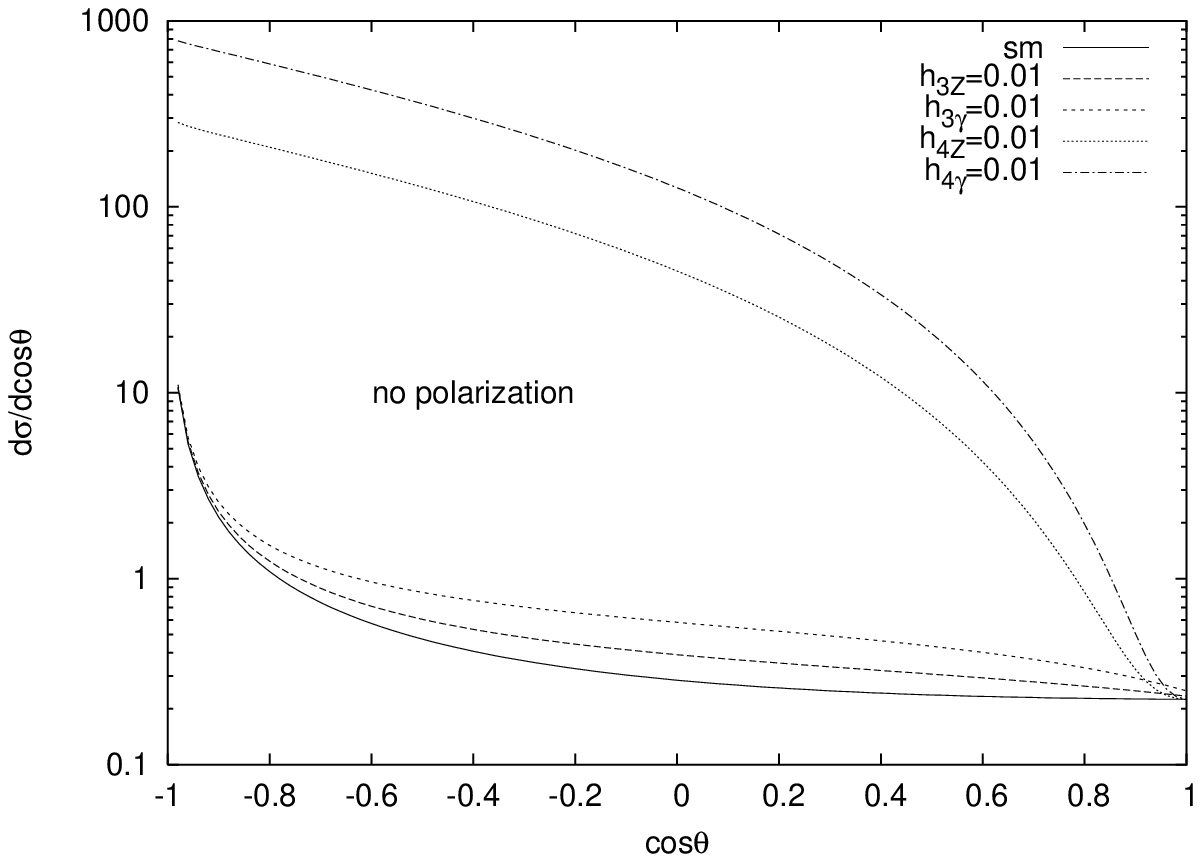}
\caption{Dependence of the angular distribution of the Z boson
on the coupling parameters of $ZZ\gamma$ and  $Z\gamma\gamma$
vertices for the unpolarized case. The unit of the cross section 
is  pb.
\label{fig6}}
\end{figure}

\begin{figure}
\includegraphics{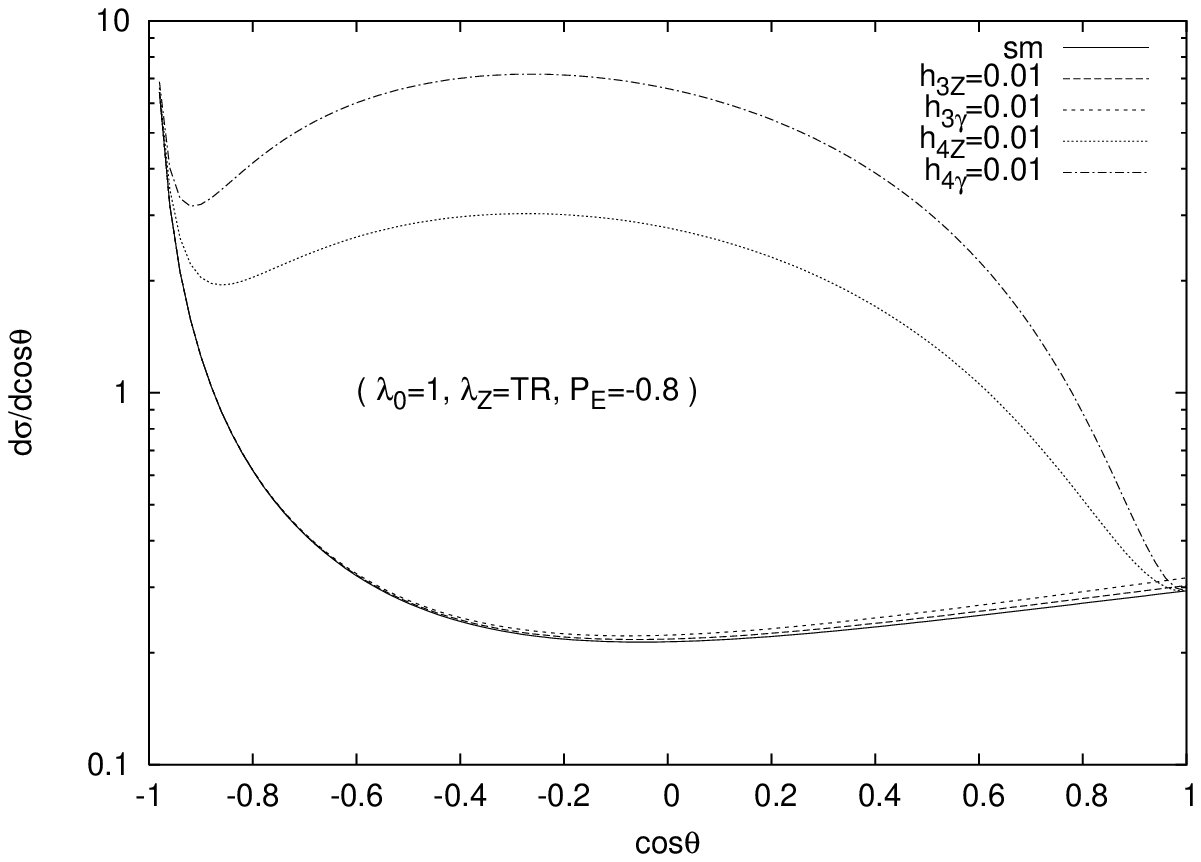}
\caption{The same as in Fig.\ref{fig6} but for the 
polarization parameters shown on the graph.
\label{fig7}}
\end{figure}

\begin{figure}
\includegraphics{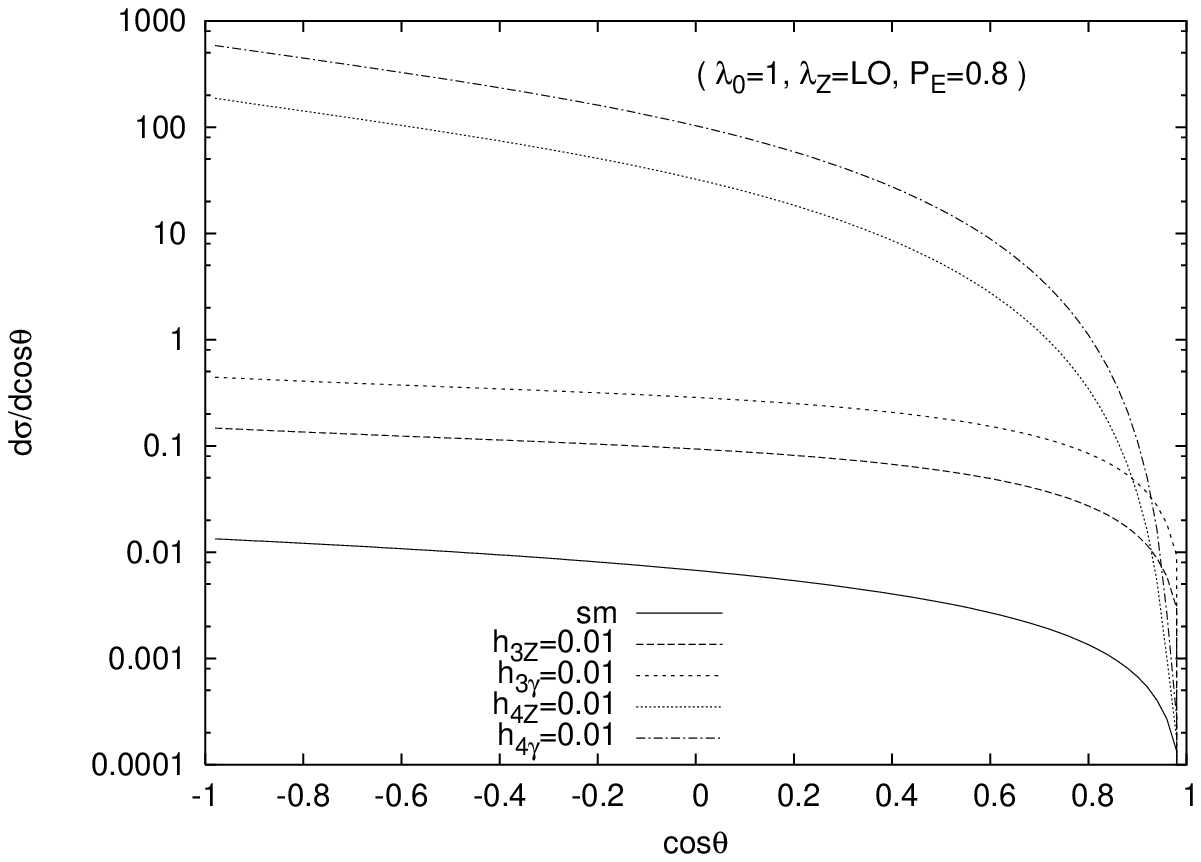}
\caption{The same as in Fig.\ref{fig6} but for the
polarization parameters shown on the graph.
\label{fig8}}
\end{figure}

It is also important to see how the anomalous couplings 
change the shape of the angular distribution of the
Z boson for the polarized and unpolarized cases. We 
use the integrated cross section formula to obtain angular 
distributions in Figs. \ref{fig6}-\ref{fig8}.  
Much larger deviations still arise from  $h_{4}^{V}$ for 
both TR and LO polarizations. The shape of the  curves 
differs for two kinds of polarizations of the Z boson. 
Additionally,  $Z\gamma\gamma$ couplings always provide the 
higher contribution to the cross section than the 
$ZZ\gamma$ couplings. Each time only one of the coupling 
parameters have been kept different from zero.
Numerical results for all polarization 
configurations will be given in the next section.

\section{Limits on the Anomalous Coupling Parameters}

In order to obtain realistic limits on the $h_{3}^{V}$
and $h_{4}^{V}$ from the LC-based $\gamma e$ collider
the number of events have been calculated 
using 
$N=A \sigma(\gamma e\to Ze) Br(Z \to \ell^{+}\ell^{-} )L_{int}$ 
for integrated luminosity $L_{int}=500 fb^{-1}$.
Here lepton channel of the Z decay  and overall acceptance 
A=0.65 has been taken into account. For total cross section 
$|\cos{\theta}|=0.9$ has been used as angular region.
The 95\% confidence level (C.L.) limits have been estimated 
using simple one parameter $\chi^{2}$ test for 
$\sqrt{s}=$0.5, 1, 1.5 TeV
from total cross section. Angular distributions
with binning procedure yield almost the same limits 
as that of total cross section.
At least 200 or more events have been considered in all 
calculations.
The limits which have been  obtained are shown in Tables 
\ref{tab1}-\ref{tab3} for the deviation of the cross section
from the standard model value without systematic error.
It should be noted that better limits are obtained for the 
polarization configuration $\lambda_{0}=-1$, $\lambda_{Z}=$LO
and $P_{e}=-1$ which leads to the order of O$(10^{-4})$ for 
$h_{3}^{Z}$ and $h_{3}^{\gamma}$, O$(10^{-6})$ for 
$h_{4}^{Z}$ and $h_{4}^{\gamma}$ at $\sqrt{s}=1.5$ TeV.
Polarization improves the limits on $h_{3}^{Z}$
and $h_{4}^{Z}$ by factors 2.5-3, $h_{3}^{\gamma}$ 
by factors 2-2.5 and $h_{4}^{\gamma}$ by factors 
2-3 depending on 
energy for the case of the polarization configuration 
given above.
In order to see the degree of energy dependence on the 
anomalous couplings let us take into account the increase 
in c.m. energy from 0.5 TeV to 1.5 TeV for the 
same polarization configuration. Then we get  
the improvements in  sensitivity limits by a factor 10 
for $h_{3}^{Z}$, by a factor 14 for 
$h_{3}^{\gamma}$, by a factor 170 for $h_{4}^{Z}$ and 
$h_{4}^{\gamma}$.

\section{Conclusion}

As the complementary collider, the $\gamma e$ mode of linear
collider with luminosity comparable to that in the collision
$e^{+}e^{-}$ probes the $ZZ\gamma$ and $ Z\gamma\gamma$
anomalous couplings with far better sensitivity than the
present colliders Fermilab Tevatron and LEP2 experiments.
Predictions from CERN LHC \cite{baur} and parental linear
$e^{+}e^{-}$ collider \cite{rosati} have been given
for coupling parameters
$f_{i}^{V}$ $(i=4,5$; $V=Z, \gamma$) due to ZZ productions.
Thus, one to one comparison with them is not possible.

In order to reach realistic results a few comments are in order. 
A reduction in luminosity is expected in the $\gamma e$
collision when compared to the basic $e^{+}e^{-}$ collision
due to scattering of laser photons. However, there are some
possibilities for increasing luminosity using electron beams
with low emittances \cite{telnov2}.

The Z boson 
polarization can be measured via lepton pair channel of the Z 
decay. Measurements of the  helicity and the angular distribution  
of the final  leptons are closely related Z boson polarization 
due to angular momentum conservation. An asymmetry measurement
with polarized Z boson can be found in Ref. \cite{sld} to 
get more information.

The results presented in this paper are based on tree level
analysis. One-loop electroweak corrections may introduce large 
logarithms as in the unpolarized total cross section. 
Furthermore, the helicity dependence of the one-loop 
amplitudes involving the exchange of Z boson is different from 
the tree-level form. This is the theoretical uncertainty 
that influences the systematic error. 

Other than the theoretical uncertainty, expected sources of systematic 
errors may result from the uncertainty on the measurement
of $\gamma e$ luminosity, helicity of incoming photons
after Compton backscattering and
uncertainty on the photon spectra \cite{gudrid}. 

For more precise results, further analysis needs to be
supplemented by observables such as the distributions
of the Z decay products with a more detailed knowledge
of the experimental conditions.

\begin{table}
\caption{Sensitivity of the $\gamma e$ collider to $ZZ\gamma$
 and $Z\gamma\gamma$ couplings at 95\% C.L. for 
$\sqrt{s}=0.5$ TeV and $L_{int}=500$ $fb^{-1}$. Only one of the 
couplings is assumed to deviate from the SM at a time.\label{tab1}}
\begin{ruledtabular}
\begin{tabular}{ccccccc}
$\lambda_{0}$  & $\lambda_{Z}$ & $P_{e}$ &$h_{3}^{Z}(10^{-2})$&
$h_{4}^{Z}(10^{-3}) $ & $h_{3}^{\gamma}(10^{-2}) $
&$h_{4}^{\gamma}(10^{-3})$ \\
\hline
0 & TR+LO & 0 &-1, 1 &-2, 2 &-0.7, 0.7 &-1, 1  \\
1 & TR & -0.8 &-3, 3 &-5, 5 &-2, 2 &-3, 3  \\
-1 &TR &-0.8 &-3, 3 &-7, 7 &-2, 2 &-4, 4 \\
1 &LO &-0.8 &-0.7, 0.7 &-1, 1 &-0.5, 0.5 &-0.7, 0.7 \\
-1 &LO &-0.8 &-0.4, 0.4 &-0.8, 0.8 &-0.3, 0.3 &-0.5, 0.5 \\
\hline
1&TR &0.8 &-3, 3 &-7, 7 &-2, 2 &-4, 4 \\
-1&TR &0.8 &-3, 3 &-6, 6 &-2, 2 &-3, 3  \\
1 &LO &0.8 &-0.5, 0.5 &-0.9, 0.9 &-0.3, 0.3 &-0.5, 0.5 \\
-1 &LO & 0.8 &-0.8, 0.8  &-1, 1 &-0.4, 0.4 &-0.6, 0.6 \\
\end{tabular}
\end{ruledtabular}
\end{table}

\begin{table}
\caption{Sensitivity of the $\gamma e$ collider to $ZZ\gamma$
and $Z\gamma\gamma$ couplings at 95\% C.L. for
$\sqrt{s}=1$ TeV and $L_{int}=500$ $fb^{-1}$. Only one of the
couplings is assumed to deviate from the SM at a time.
\label{tab2}}
\begin{ruledtabular}
\begin{tabular}{ccccccc}
$\lambda_{0}$  & $\lambda_{Z}$ & $P_{e}$ &$h_{3}^{Z}(10^{-3})$&
$h_{4}^{Z}(10^{-5}) $ & $h_{3}^{\gamma}(10^{-3}) $
&$h_{4}^{\gamma}(10^{-5})$ \\
\hline
 0 & TR+LO & 0 &-2, 2 &-9, 9 &-1, 1 &-5, 5  \\
 1 & TR & -0.8 &-10, 10 &-50, 50 &-7, 7 &-30, 30  \\
 -1 &TR &-0.8 &-10, 10 &-60, 60 &-8, 8 &-40, 40 \\
 1 &LO &-0.8 &-1, 1 &-5, 5 &-0.8, 0.8 &-3, 3 \\
 -1 &LO &-0.8 &-0.7, 0.7 &-3, 3 &-0.5, 0.5 &-2, 2 \\
\hline
 1&TR &0.8 &-10, 10 &-70, 70 &-7, 7 &-40, 40 \\
 -1&TR &0.8 &-10, 10 &-50, 50 &-6, 6 &-30, 30  \\
 1 &LO &0.8 &-0.9, 0.9 &-4, 4 &-0.5, 0.5 &-2, 2 \\
 -1 &LO & 0.8 &-1, 1  &-5, 5 &-0.8, 0.8 &-3, 3 \\
\end{tabular}
\end{ruledtabular}
\end{table}

\begin{table}
\caption{Sensitivity of the $\gamma e$ collider to $ZZ\gamma$
and $Z\gamma\gamma$ couplings at 95\% C.L. for
$\sqrt{s}=1.5$ TeV and $L_{int}=500$ $fb^{-1}$. Only one of the
couplings is assumed to deviate from the SM at a time.
\label{tab3}}
\begin{ruledtabular}
\begin{tabular}{ccccccc}
$\lambda_{0}$  & $\lambda_{Z}$ & $P_{e}$ &$h_{3}^{Z}(10^{-4})$&
$h_{4}^{Z}(10^{-5}) $ & $h_{3}^{\gamma}(10^{-4}) $
&$h_{4}^{\gamma}(10^{-5})$ \\
\hline
0 & TR+LO & 0 &-9, 9 &-1, 1 &-5, 5 &-0.9, 0.9  \\
1 & TR & -0.8 &-60, 60 &-10, 10 &-40, 40 &-8, 8  \\
-1 &TR &-0.8 &-70, 70 &-20, 20 &-40, 40 &-10, 10 \\
1 &LO &-0.8 &-5, 5 &-0.7, 0.7 &-3, 3 &-0.5, 0.5 \\
-1 &LO &-0.8 &-3, 3 &-0.5, 0.5 &-2, 2 &-0.3, 0.3 \\
\hline
1&TR &0.8 &-70, 70 &-20, 20 &-40, 40 &-10, 10 \\
-1&TR &0.8 &-60, 60 &-10, 10 &-30, 30 &-7, 7  \\
1 &LO &0.8 &-3, 3 &-0.6, 0.6 &-2, 2 &-0.3, 0.3 \\
-1 &LO & 0.8 &-5, 5  &-0.8, 0.8 &-3, 3 &-0.4, 0.4 \\
\end{tabular}
\end{ruledtabular}
\end{table}

\turnpage

\appendix*
\section{Helicity Amplitudes}
There are four Feynman diagrams for the process 
$\gamma e\to Ze$ if one includes  $ZZ\gamma$, $Z\gamma\gamma$
vertices.
Helicity amplitudes $M_{1}$ and $M_{4}$ are responsible
for the diagrams concerning  $ZZ\gamma$ and $Z\gamma\gamma$ 
interactions arising from t channel Z or $\gamma$ exchanges. 
$M_{2}$ and $M_{3}$ are  standard model 
contribution of the u and s channel of the process.
The parameters of the helicity amplitudes 
$M(\lambda_{\gamma}, \sigma_{e}; \lambda_{Z}, \sigma_{e}^{\prime})$
are helicities of incoming photon and electron,  
outgoing Z boson  and electron. The values they take are given
by :
\begin{eqnarray}
\lambda_{\gamma} : +, - ~~,~~ \sigma_{e}: L, R~~,~~
\lambda_{Z} : +, -, 0 ~~,~~ \sigma_{e}^{\prime}: L, R 
\end{eqnarray}
Here L and R stand for left and right.
Helicity amplitudes we have obtained for each Feynman 
diagram in the c.m. frame of  $\gamma e$
can be written as follows:

\begin{eqnarray}
M(\lambda_{\gamma}, \sigma_{e}; \lambda_{Z}, \sigma_{e}^{\prime})
&&=M_{1}(\lambda_{\gamma}, \sigma_{e}; \lambda_{Z}, \sigma_{e}^{\prime})
+M_{2}(\lambda_{\gamma}, \sigma_{e}; \lambda_{Z}, \sigma_{e}^{\prime})
\nonumber \\
&&+M_{3}(\lambda_{\gamma}, \sigma_{e}; \lambda_{Z}, \sigma_{e}^{\prime})
+M_{4}(\lambda_{\gamma}, \sigma_{e}; \lambda_{Z}, \sigma_{e}^{\prime})
\end{eqnarray}

\begin{eqnarray}
M_{1}(+L;+L)&&=C_{1}^{L}\left\{h_{3}^{Z} \left [iE_{1}(\sin{\theta}
\sin{\frac{\theta}{2}}-2\cos{\frac{\theta}{2}})\right ]
+\frac{h_{4}^{Z}}{M_{Z}^{2}}\left[-iE_{1}^{2}p_{3}
(1+\cos{\theta})\sin{\frac{\theta}{2}}\sin{\theta}\right]\right\}
\\
M_{1}(+L;-L)&&=C_{1}^{L}\left\{h_{3}^{Z} \left[-iE_{1}
\sin{\frac{\theta}{2}}\sin{\theta} \right]
+\frac{h_{4}^{Z}}{M_{Z}^{2}}\left[ip_{3}E_{1}^{2}
\cos{\frac{\theta}{2}}\sin^{2}{\theta}\right]\right\}
\\
M_{1}(+L;0L)&&=C_{1}^{L}\frac{\sqrt{2}}{M_{Z}} \left\{h_{3}^{Z} 
\left[iE_{1}E_{3}(1+\cos{\theta})
\sin{\frac{\theta}{2}}\right]
+\frac{h_{4}^{Z}}{M_{Z}^{2}} 
\left[ip_{3}E_{1}^{2}\sin{\frac{\theta}{2}}
(p_{3}-E_{3}\cos{\theta})(1+\cos{\theta})\right]\right\}
\\
M_{1}(-L;+L)&&=C_{1}^{L}\left\{
\frac{h_{4}^{Z}}{M_{Z}^{2}}\left[-iE_{1}^{2}(p_{3}+E_{3})
\sin{\frac{\theta}{2}}\sin{\theta} \right]\right\}
\\
M_{1}(-L;0L)&&=C_{1}^{L}\frac{\sqrt{2}}{M_{Z}}
\left\{h_{3}^{Z} \left[iE_{1}(p_{3}+E_{3})\sin{\frac{\theta}{2}}
\right]
+\frac{h_{4}^{Z}}{M_{Z}^{2}}\left[iE_{1}^{2}
\sin{\frac{\theta}{2}}(p_{3}^{2}+E_{3}p_{3}(1-\cos{\theta})
-E_{3}^{2}\cos{\theta})\right]\right\}
\\
M_{1}(-L;-L)&&=C_{1}^{L}\left\{h_{3}^{Z} \left[2iE_{1}
\cos{\frac{\theta}{2}}\right]
+\frac{h_{4}^{Z}}{M_{Z}^{2}}\left[iE_{1}^{2}(p_{3}+E_{3})
\sin{\frac{\theta}{2}}\sin{\theta}
\right]\right\}
\\
M_{1}(+R;+R)&&=C_{1}^{R}\left\{h_{3}^{Z} 
\left[-2iE_{1}\cos{\frac{\theta}{2}} \right]
+\frac{h_{4}^{Z}}{M_{Z}^{2}}
\left[-iE_{1}^{2}(p_{3}+E_{3})\sin{\frac{\theta}{2}}\sin{\theta}
\right]\right\}
\\
M_{1}(+R;-R)&&=C_{1}^{R}\left\{ 
\frac{h_{4}^{Z}}{M_{Z}^{2}}\left[iE_{1}^{2}(p_{3}+E_{3})
\sin{\frac{\theta}{2}}\sin{\theta}
\right]\right\}
\\
M_{1}(+R;0R)&&=C_{1}^{R}\frac{\sqrt{2}}{M_{Z}}
\left\{h_{3}^{Z} \left[iE_{1}(p_{3}+E_{3})
\sin{\frac{\theta}{2}}\right]
+\frac{h_{4}^{Z}}{M_{Z}^{2}}
\left[iE_{1}^{2}\sin{\frac{\theta}{2}}
(p_{3}^{2}+E_{3}p_{3}(1-\cos{\theta})
-E_{3}^{2}\cos{\theta})\right]\right\}
\\
M_{1}(-R;+R)&&=C_{1}^{R}\left\{h_{3}^{Z} 
\left[iE_{1}\sin{\frac{\theta}{2}}\sin{\theta} \right]
+\frac{h_{4}^{Z}}{M_{Z}^{2}}
\left[-iE_{1}^{2}p_{3}(1+\cos{\theta})
\sin{\frac{\theta}{2}}\sin{\theta}\right]\right\}
\\
M_{1}(-R;-R)&&=C_{1}^{R}\left\{h_{3}^{Z} 
\left[iE_{1}\cos{\frac{\theta}{2}}(1+\cos{\theta})\right]
+\frac{h_{4}^{Z}}{M_{Z}^{2}}\left[iE_{1}^{2}p_{3}
(1+\cos{\theta})\sin{\frac{\theta}{2}}\sin{\theta}
\right]\right\}
\\
M_{1}(-R;0R)&&=C_{1}^{R}\frac{\sqrt{2}}{M_{Z}}
\left\{h_{3}^{Z} \left[iE_{1}E_{3}(1+\cos{\theta})
\sin{\frac{\theta}{2}}\right]
+\frac{h_{4}^{Z}}{M_{Z}^{2}}\left[
iE_{1}^{2}p_{3}\sin{\frac{\theta}{2}}(1+\cos{\theta})
(p_{3}-E_{3}\cos{\theta}) \right]\right\}
\end{eqnarray}
where
\begin{eqnarray}
C_{1}^{L}=-\frac{2g_{e}g_{L}\sqrt{E_{2}E_{4}}}{M_{Z}^{2}}
~, ~~~
C_{1}^{R}=-\frac{2g_{e}g_{R}\sqrt{E_{2}E_{4}}}{M_{Z}^{2}}
\end{eqnarray}
with 
\begin{eqnarray}
g_{L}=\frac{g_{Z}}{2}(C_{V}+C_{A})~, ~~~ 
g_{R}=\frac{g_{Z}}{2}(C_{V}-C_{A})
\\
C_{V}=2\sin^{2}{\theta_{W}}-\frac{1}{2}~, ~~~C_{A}=-\frac{1}{2}
\\
g_{Z}=\frac{g_{e}}{\sin{\theta_{W}}\cos{\theta_{W}}}
~, ~~~g_{e}^{2}=4\pi\alpha
\end{eqnarray}
Here $E_{1}$: energy of incoming photon, $E_{2}$: energy of 
incoming electron, $E_{3}$: energy of outgoing Z boson,
$p_{3}=|\vec{p}_{3}|$: magnitude of the outgoing Z boson 
momentum, $E_{4}$: energy of outgoing electron.

\begin{eqnarray}
M_{2}(+L;+L)&&=C_{2}^{L}\left\{-(p_{3}+E_{3})
(1+\cos{\theta})\cos{\frac{\theta}{2}}\right\}
\\
M_{2}(+L;-L)&&=C_{2}^{L}\left\{(p_{3}-E_{3})
(1-\cos{\theta})\cos{\frac{\theta}{2}}\right\}
\\
M_{2}(+L;0L)&&=C_{2}^{L}\left\{\sqrt{2}M_{Z}
\sin{\theta}\cos{\frac{\theta}{2}}\right\}
\\
M_{2}(-L;+L)&&=C_{2}^{L}\left\{(2E_{1}-p_{3}-E_{3})
\sin{\theta}\sin{\frac{\theta}{2}}\right\}
\\
M_{2}(-L;-L)&&=C_{2}^{L}\left\{(E_{3}-p_{3}-2E_{1})
\sin{\theta}\sin{\frac{\theta}{2}}\right\}
\\
M_{2}(-L;0L)&&=C_{2}^{L}\frac{\sqrt{2}}{M_{Z}}
\left\{ \left[(2E_{1}E_{3}-M_{Z}^{2})\cos{\theta}
+2E_{1}p_{3}\right]\sin{\frac{\theta}{2}}\right\}
\\
M_{2}(+R;+R)&&=C_{2}^{R}\left\{(E_{3}-p_{3}-2E_{1})
\sin{\theta}\sin{\frac{\theta}{2}}\right\}
\\
M_{2}(+R;-R)&&=C_{2}^{R}\left\{(2E_{1}-E_{3}-p_{3})
\sin{\theta}\sin{\frac{\theta}{2}}\right\}
\\
M_{2}(+R;0R)&&=C_{2}^{R}\frac{\sqrt{2}}{M_{Z}}
\left\{ \left[(M_{Z}^{2}-2E_{1}E_{3})\cos{\theta}
-2E_{1}p_{3}\right]\sin{\frac{\theta}{2}}\right\}
\\
M_{2}(-R;+R)&&=C_{2}^{R}\left\{(p_{3}-E_{3})
(1-\cos{\theta})\cos{\frac{\theta}{2}}\right\}
\\
M_{2}(-R;-R)&&=C_{2}^{R}\left\{-(p_{3}+E_{3})
(1+\cos{\theta})\cos{\frac{\theta}{2}}\right\}
\\
M_{2}(-R;0R)&&=C_{2}^{R}\left\{-\sqrt{2}M_{Z}
\sin{\theta}\cos{\frac{\theta}{2}}\right\}
\end{eqnarray}
where
\begin{eqnarray}
&&C_{2}^{L}=\frac{2Q_{e}g_{e}g_{L}\sqrt{E_{2}E_{4}}}
{\hat{u}-M_{e}^{2}}~,~~~
C_{2}^{R}=\frac{2Q_{e}g_{e}g_{R}\sqrt{E_{2}E_{4}}}
{\hat{u}-M_{e}^{2}} \\
&&\hat{u}=M_{Z}^{2}-2E_{2}(E_{3}+p_{3}\cos{\theta})
~,~~~Q_{e}=-1
\end{eqnarray}

\begin{eqnarray}
M_{3}(+L;+L)&&=0~,~~~M_{3}(+L;-L)=0
\\
M_{3}(+L;0L)&&=0~,~~~M_{3}(-L;+L)=0
\\
M_{3}(-L;-L)&&=C_{3}^{L}\left\{4E_{1}\cos{\frac{\theta}{2}} 
\right\}
\\
M_{3}(-L;0L)&&=C_{3}^{L}\frac{\sqrt{2}}{M_{Z}}\left\{
2E_{1}(p_{3}+E_{3})\sin{\frac{\theta}{2}}\right\}
\\
M_{3}(+R;+R)&&=C_{3}^{R}\left\{4E_{1}\cos{\frac{\theta}{2}}
\right\}\\
M_{3}(+R;-R)&&=0 \\
M_{3}(+R;0R)&&=C_{3}^{R}\frac{\sqrt{2}}{M_{Z}}\left\{
-2E_{1}(p_{3}+E_{3})\sin{\frac{\theta}{2}}\right\}
\\
M_{3}(-R;+R)&&=0~,~~~M_{3}(-R;-R)=0 \\
M_{3}(-R;0R)&&=0
\end{eqnarray}
where
\begin{eqnarray}
&&C_{3}^{L}=\frac{2Q_{e}g_{e}g_{L}\sqrt{E_{2}E_{4}}}
{\hat{s}-M_{e}^{2}}~,~~~
C_{3}^{R}=\frac{2Q_{e}g_{e}g_{R}\sqrt{E_{2}E_{4}}}
{\hat{s}-M_{e}^{2}} 
\end{eqnarray}

\begin{eqnarray}
&&M_{4}(\lambda_{\gamma}, L; \lambda_{Z},L)=
C_{4}^{L}\left(\frac{1}{C_{1}^{L}}
M_{1}(\lambda_{\gamma}, L; \lambda_{Z},L)\right)
\\
&&M_{4}(\lambda_{\gamma}, R; \lambda_{Z},R)=
C_{4}^{R}\left(\frac{1}{C_{1}^{R}}
M_{1}(\lambda_{\gamma}, R; \lambda_{Z},R)\right)
\\
&&h_{3}^{Z}\to h_{3}^{\gamma}~, 
~~~h_{4}^{Z}\to h_{4}^{\gamma}
\end{eqnarray}
where
\begin{eqnarray}
C_{4}^{L}=C_{4}^{R}=
-\frac{2Q_{e}g_{e}^{2}\sqrt{E_{2}E_{4}}}{M_{Z}^{2}}
\end{eqnarray}

\end{document}